\documentclass[%
 reprint,
nofootinbib,
 amsmath,amssymb,
 aps,
]{revtex4-2}

\usepackage{graphicx}
\usepackage{dcolumn}
\usepackage{bm}
\usepackage{hyperref}
\usepackage{ulem}
\usepackage{xcolor}

\definecolor{deletepurple}{RGB}{140,60,160}

\begin{document}

\preprint{APS/123-QED}

\title{Effects of the $a_0(980)$ Meson on Finite Nuclei in a Parity Doublet Model}

\author{Yuk Kei Kong$^{a,b}$}
 \email{hv.14j.8236@s.thers.ac.jp}
\author{Youngman Kim$^c$}%
 \email{ykim@ibs.re.kr}
\author{Masayasu Harada$^{d,a,e}$}
 \email{harada.masayasu@nagoya-u.jp}

\affiliation{$^a$Department of Physics, Nagoya University, Nagoya, 464-8602, Japan}
\affiliation{$^b$TOPTIER, Institute for Rare Isotope Science, Institute for Basic Science, Daejeon 34000, Republic of Korea
}%
\affiliation{$^c$Center for Exotic Nuclear Studies, Institute for Basic Science, Daejeon 34126, Republic of Korea
}%
\affiliation{$^d$Kobayashi-Maskawa Institute for the Origin of Particles and the Universe, Nagoya University, Nagoya, 464-8602, Japan}
\affiliation{$^e$Advanced Science Research Center, Japan Atomic Energy Agency, Tokai, 319-1195, Japan}

\date{\today}

\begin{abstract}
We investigate nuclear structure within the parity doublet model (PDM), with particular emphasis on the role of the isovector-scalar $a_0(980)$ meson. 
We adopt the models with and without the $a_0$ meson and fit their parameters to nuclear matter saturation and ground state properties of selected nuclei, subject to constraints from the vacuum. 
The resulting parameter sets reproduce the nuclear matter saturation properties and provide a good description of binding energies and charge radii of selected nuclei.
Our fits show that a large value of the chiral invariant mass of nucleon is preferred by the fitting: the best fitted value is $m_0 =  689 \pm 71 $\,MeV with $a_0$ and $m_0 = 689 \pm 69 $\,MeV without $a_0$.
We find that the effects of the $a_0$ meson on the binding energies and charge radii are negligibly small. 
It is difficult to clarify the effect of $a_0$ meson to neutron skin thickness of Sn, Pb, and Ca isotopes, due to the large uncertainty of the predictions.
We note that our prediction of $^{208} \text{Pb}$ is consistent with the measurement by PREX-II collaboration, while that of $^{48} \text{Ca}$ cannot explain the CREX result.

%
\end{abstract}

\maketitle


\section{\label{sec:intro}Introduction}

Chiral symmetry and its spontaneous breaking are the essential properties in low energy dynamics of quantum chromodyanmics (QCD). 
In particular, the spontaneous chiral symmetry breaking 
(S$\chi$SB) generates a part of hadron masses.

The parity doublet model (PDM)~\cite{PhysRevD.39.2805,10.1143/PTP.106.873} provides a distinctive realization of chiral symmetry in the baryon sector. In the PDM, an excited nucleon such as $N(1535)$ is regarded as the chiral partner to ground-state nucleon $N(939)$, and the mass difference of the chiral partners is generated by the 
S$\chi$SB.
When the chiral symmetry is restored, their masses are degenerated to the chiral invariant mass $m_0$.
The existence of $m_0$ is supported by the lattice QCD simulations~\cite{PhysRevD.92.014503,2017JHEP...06..034A,PhysRevD.99.074503} and QCD sum rule calculations~\cite{PhysRevD.103.L051501,PhysRevD.105.014014,sym15040799}.

Various attempts have been made to constrain $m_0$ from hadronic, nuclear, and astrophysical observables. Vacuum analyses based on the properties and decays of the negative-parity nucleon provide constraints that depend on the assumed interactions and baryon representations~\cite{PhysRevD.39.2805,PhysRevD.57.4124,10.1143/PTP.106.873,PhysRevD.82.014004,PhysRevD.85.054022,PhysRevD.99.034012}. The PDM has also been extensively applied to study dense nuclear matter and neutron stars~\cite{Hatsuda:1988mv, Zschiesche:2006zj, Dexheimer:2007tn, Dexheimer:2008cv, Sasaki:2010bp, Sasaki:2011ff,%
Gallas:2011qp, Paeng:2011hy,%
Steinheimer:2011ea,Dexheimer:2012eu, Paeng:2013xya,Benic:2015pia,Motohiro:2015taa,%
Mukherjee:2016nhb,Suenaga:2017wbb,Takeda:2017mrm,Mukherjee:2017jzi,Paeng:2017qvp,%
Marczenko:2017huu,Abuki:2018ijb,Marczenko:2018jui,Marczenko:2019trv,Yamazaki:2019tuo,%
Harada:2019oaq,Marczenko:2020jma,Harada:2020etl,%
PhysRevC.103.045205,Marczenko:2021uaj,PhysRevC.104.065201,Marczenko:2022hyt,%
PhysRevC.106.065205,Minamikawa:2023key,Marczenko:2023ohi,PhysRevC.108.055206,%
universe11100345,xwkv-s8lg,nlhh-xqjp,%
SciChinaPhysMechAstron6920263232011, PhysRevC.109.065807,sym16091238%
}.
The sensitivity of the resulting NS properties to $m_0$, together with observational constraints on NS such as masses and radii, has been used to constrain the chiral invariant mass
(see, e.g., Refs.~\cite{PhysRevC.103.045205,PhysRevC.104.065201,PhysRevC.106.065205,Minamikawa:2023key,universe11100345,PhysRevC.108.055206}).
More recently, the nucleon gravitational D-form factor~\cite{KAWAGUCHI2026140400} and electromagnetic form factor~\cite{Harada:2026dyj}
are studied as additional probes of the chiral invariant mass. 
Taken together, these studies suggest a chiral invariant mass in the approximate range $m_0 \sim 400$–$900$ MeV, indicating that a substantial portion of the nucleon mass may be independent of the S$\chi$SB.

On the other hand, study of nuclear structure provides us a complementary way to examine the realization of chiral symmetry at densities around and below saturation density. 
It also allows us to constrain $m_0$ using nuclear ground-state data.
The first study based on the parity doublet model in Ref.~\cite{YKimNSPDM2023} found that $m_0 \simeq 700$ MeV is preferred by the binding energies and charge radii of selected nuclei.
In that study, the model parameters are determined from vacuum and nuclear saturation properties for several choices of $m_0$, and the predicted binding energies and charge radii of selected nuclei are compared with experimental data.
This finding motivates a systematic determination of $m_0$ by incorporating finite-nucleus observables directly into the parameter calibration and assessing the associated uncertainties.

In recent years, an increasing number of researchers have become interested in neutron-rich systems for understanding the isovector sector of nuclear interactions, which is essential for describing asymmetric matter such as neutron rich nuclei and neutron star matter.
In particular, there are many works to investigate the role of the isovector scalar $a_0(980)$ meson based on conventional Walecka-type RMF models~\cite{Kubis_1997,https://doi.org/10.48550/arxiv.astro-ph/9802303,Miyatsu_2022,Li_2022,https://doi.org/10.48550/arxiv.2209.02861,Thakur_2022,Liu_2005,PhysRevC.80.025806,Gaitanos_2004,PhysRevC.67.015203,PhysRevC.65.045201}, density-dependent RMF approaches~\cite{PhysRevC.90.055801,PhysRevC.84.054309}, 
and PDM~\cite{PhysRevC.108.055206,sym16091238,Yukkei:2024jxa,xwkv-s8lg,universe11100345} in the context of asymmetric nuclear matter and neutron-star matter.
These studies have shown that the $a_0$ meson significantly modifies the isovector properties such as the equation of state of neutron-rich matter and the density dependence of the symmetry energy.
It is therefore natural to extend these investigations to finite nuclei and examine how the $a_0$ meson influences their properties.

In this work, we study the nuclear structure  within the PDM, focusing on the effects of the isovector scalar $a_0(980)$ meson. We consider models with and without the $a_0$ meson, $a_0$ model and no-$a_0$ model, and determine the model parameters by fitting them to nuclear matter saturation properties and ground-state properties of selected  nuclei. 
Then, we investigate the  effect of the $a_0$ meson  to
the saturation properties of nuclear matter, the density dependence of the symmetry energy, binding energies, charge radii, and neutron skin thicknesses. 
We also compare the predicted neutron skin thicknesses of $^{48}\mathrm{Ca}$ and $^{208}\mathrm{Pb}$ with the CREX~\cite{PhysRevLett.129.042501} and PREX-II~\cite{PhysRevLett.126.172502}
measurements.

This paper is organized as follows. In Sec.~\ref{sec:TF}, we introduce the PDM and 
present the relativistic Hartree equations.
Section~\ref{sec:level_i} describes the procedure used to determine the model parameters. The  nuclear matter saturation properties calculated from the present models are discussed in Sec.~\ref{sec:SP}. In Sec.~\ref{sec:NP}, we present the results for nuclei, including the binding energies, charge radii, and neutron skin thicknesses. We also discuss the results for the neutron skin thickness in relation to the CREX and PREX-II measurements. 
Finally, a summary and discussion are given in Sec.~\ref{sec:NM}.

\section{\label{sec:TF}Formulism}

\subsection{Parity Doublet Model}

We employ the parity doublet model including the iso-vector scalar meson $a_0(980)$ constructed in Refs.~\cite{PhysRevC.108.055206,sym16091238}. The model is based on the $SU(2)_L \times SU(2)_R \times U(1)_A$ symmetry, with the $U(1)_A$ symmetry explicitly broken by the anomaly term. In the model, the scalar meson fields and their chiral partner $\sigma, \pi, a_0, \eta$, and the vector meson fields $\omega, \rho$ are introduced based on the hidden local symmetry~\cite{Bando:1987br,Harada:2003jx}. In addition, the electromagnetic field $A_\mu$ is introduced to account for the Coulomb interaction among protons in finite nuclei. 
In the Hartree approximation, the meson and electromagnetic field operators are replaced by their expectation values. For a static, time-reversal-invariant even-even nucleus, these mean fields take the following forms:
\begin{align}
\sigma(x)
&\rightarrow
\langle \sigma(x) \rangle
\equiv \sigma(\mathbf{x}),\nonumber
\\
\pi^a(x)
&\rightarrow 0,\nonumber
\\
a_0^a(x)
&\rightarrow
\langle a_0^a(x) \rangle
\equiv \delta^{a3}a(\mathbf{x}),\nonumber
\\
\eta(x)
&\rightarrow 0,\nonumber
\\
\omega^\mu(x)
&\rightarrow
\langle \omega^\mu(x) \rangle
\equiv\delta^{\mu0}\omega(\mathbf{x}),\nonumber
\\
\rho^{a\mu}(x)
&\rightarrow
\langle \rho^{a\mu}(x) \rangle
\equiv\delta^{a3}\delta^{\mu0}\rho(\mathbf{x}),\nonumber
\\
A^\mu(x)
&\rightarrow
\langle A^\mu(x) \rangle
\equiv\delta^{\mu0}A_0(\mathbf{x}).
\end{align}
The pseudoscalar mean fields vanish owing to parity conservation, while only the third isospin components of the isovector fields remain. Time-reversal invariance implies that the spatial components of the vector fields vanish. Therefore, the meson field $M = [\sigma + i\vec{\pi} \cdot \vec{\tau}] -   [ \vec{a_0} \cdot \vec{\tau} + i\eta] $ becomes
\begin{equation}
M \rightarrow \left\langle M \right\rangle = \begin{pmatrix}
\sigma - a & 0 \\ 0 & \sigma + a
\end{pmatrix}
\ .
\end{equation}
The vacuum expectation value of $M$ is defined as 
\begin{equation}
\begin{aligned}
  \langle 0 | M | 0 \rangle = \begin{pmatrix}
\sigma_0 & 0\\
0 & \sigma_0
\end{pmatrix},
\end{aligned}
\end{equation} 
where $\sigma_0 = \langle 0|\sigma|0 \rangle$ is the VEV of the $\sigma$ field. In vacuum, $\sigma_0$ is equal to the pion decay constant $f_\pi = 92.4 $ MeV.

The PDM Larangian is divided into four parts:
\begin{equation}
    \mathcal{L} = \mathcal{L}_N + \mathcal{L}_M  +  \mathcal{L}_{V} +  \mathcal{L}_{EM}\ , \label{PDM-L1}
\end{equation} 
where 
$\mathcal{L}_N$, $\mathcal{L}_M$, $\mathcal{L}_{V}$ and  $\mathcal{L}_{EM}$ denote
the nucleon Lagrangian, 
the scalar meson Lagrangian,
the vector meson Lagrangian,
and the electromagnetic Lagrangian, 
respectively.

The baryonic Lagrangian $\mathcal{L}_N$ based on the parity doubling structure \cite{PhysRevD.39.2805,10.1143/PTP.106.873} is given by 
\begin{align}
\mathcal{L}_{N}
=
\sum_{\substack{\alpha=\pm\\ j=p,n}}
\bar{N}_{\alpha j}
\left[
i\gamma_{\mu}\partial^{\mu}
-
m_{\alpha j}^{*}
\right]
N_{\alpha j},
\end{align}
where $\alpha = \pm$ denotes the parity of the nucleon,
and $j = p, n$ labels proton ($j = +1$) and neutron ($j = -1$). $m_{\alpha j}^{*}$ is the effective nucleon mass~\cite{PhysRevC.108.055206,sym16091238} given by~\footnote{
Here $\sigma$ and $a$ are $\sigma(\mathbf{x})$ and $a(\mathbf{x})$. In what follows, we omit the function arguments.
}
\begin{equation}
\begin{aligned}
  m^*_{\alpha j} = \frac{1}{2} \bigg[ \sqrt{(g_1+g_2)^2(\sigma - ja)^2 + 4m_0^2} + \alpha(g_1 - g_2)(\sigma - ja) \bigg]\ .
\end{aligned}
\label{maj}
\end{equation}
The meson Lagrangian $\mathcal{L}_M$ in the mean field approximation is given by
\begin{align}
\mathcal{L}_M
&=
\frac{1}{4}
\operatorname{tr}
\left[
\partial_\mu M
\partial^\mu M^\dagger
\right]
-
V_M
\nonumber\\
&=
\frac{1}{2} \partial_\mu\sigma\, \partial^\mu\sigma
+
\frac{1}{2}
\partial_\mu a\,
\partial^\mu a
-
V_M.
\end{align}
where the potential $V_M$ 
is expressed as
\begin{equation}
\begin{aligned}
V_M = & - \frac{\bar{\mu}^2_{\sigma}}{2} \sigma^2 - \frac{\bar{\mu}^2_{a}}{2} a^2 + \frac{\lambda_4}{4} (\sigma^4 + a^4 ) + \frac{\gamma_4}{2} \sigma^2 a^2 \\ 
& {} - \frac{\lambda_6}{6} (\sigma^6 +15\sigma^2a^4 + 15\sigma^4a^2 +a^6 ) +\lambda_6^{'}(\sigma^2a^4 + \sigma^4a^2) \\ 
& {} - m^2_{\pi} f_{\pi} \sigma \ .
\end{aligned}
\label{VMa}
\end{equation}
The vector meson Lagrangian $\mathcal{L}_{V}$ in the mean field approximation is given by
\begin{equation}
\begin{aligned}
\mathcal{L}_{V} = & - g_{\omega NN} \sum_{\alpha j} \bar{N}_{\alpha j} \gamma^0 \omega N_{\alpha j} - g_{\rho NN} \sum_{\alpha j} \bar{N}_{\alpha j} \gamma^0 \frac{\tau_3}{2} \rho N_{\alpha j} \\
& -\frac{1}{4}\omega_{\mu\nu}\omega^{\mu\nu} -\frac{1}{4}\rho_{\mu\nu}\rho^{\mu\nu} \\
& + \frac{1}{2} m^2_{\omega} \omega^2 + \frac{1}{2} m^2_{\rho} \rho^2 + \lambda_{\omega \rho} g_{\omega NN}^2 g_{\rho NN}^2 \omega^2 \rho^2\ .
\end{aligned}
\end{equation} 
where $\omega_{\mu\nu}=\partial_\mu\omega \delta_{\nu0}-\partial_\nu\omega \delta_{\mu0} $ and $\rho_{\mu\nu}=\partial_\mu\rho\delta_{\nu0}-\partial_\nu\rho\delta_{\mu0}$.
The electromagnetic field interaction is described by the the electromagnetic Lagrangian
\begin{equation}
\mathcal{L}_{EM}= - e \sum_{\alpha} \bar{N}_{\alpha p} \gamma^0 A_0 N_{\alpha p} -\frac{1}{4}F_{\mu\nu}F^{\mu\nu},
\end{equation} 
where $F_{\mu\nu} = \partial_\mu A_0 \delta_{\nu0} -\partial_\nu A_0 \delta_{\mu0}$ denotes the electromagnetic field tensor.

\subsection{Hamiltonian and Equation of Motion}

The corresponding Hamiltonian is obtained by performing the Legendre transformation to the Lagrangian as
\begin{widetext}
\begin{align}
\mathcal{H}
={}&
\bar{N}
\left(
-i\gamma^{i}\partial_{i}
+m_{N}^{*}
\right)N
+g_{\omega NN}\omega N^{\dagger}N
+g_{\rho NN}\rho
N^{\dagger}\frac{\tau_{3}}{2}N
+eA_{0}
N^{\dagger}\frac{1-\tau_{3}}{2}N
\nonumber \\ 
&-\frac{1}{2}
(\partial^{i}\sigma)(\partial_{i}\sigma)
+\frac{1}{2}
(\partial^{i}\omega)(\partial_{i}\omega)
+\frac{1}{2}
(\partial^{i}\rho)(\partial_{i}\rho)
+\frac{1}{2}
(\partial^{i}A_{0})(\partial_{i}A_{0})
\nonumber \\
& + V_M
\nonumber \\
&-\frac{1}{2}m_{\omega}^{2}\omega^{2}
-\frac{1}{2}m_{\rho}^{2}\rho^{2} -  \lambda_{\omega \rho} g_{\omega NN}^2 g_{\rho NN}^2 \omega^2 \rho^2 \ ,
\label{eq:Hamiltonian_density}
\end{align}
\end{widetext}
where spatial index takes $i=1,2,3$ and we take the temporal derivatives as zero. We note that in this work, the negative-parity nucleons are neglected in the finite nuclei calculations because their Fermi sea is formed only at densities much higher than the nuclear saturation density.

By performing variation respect to the meson fields, the equations of motion (EoMs) are obtained as
\begin{align}
-\nabla^2 \sigma(\mathbf{r})
&=
-\sum_{j}
\bar{N}_{+j}(\mathbf{r})
N_{+j}(\mathbf{r})
\frac{\partial m^{*}_{+j}}{\partial\sigma}
\nonumber\\
&\quad
-
\frac{\partial V_M}{\partial\sigma},
\\
-\nabla^2 a(\mathbf{x})
&=
-\sum_{j}
\bar{N}_{+j}(\mathbf{x})
N_{+j}(\mathbf{x})
\frac{\partial m^{*}_{+j}}{\partial a}
\nonumber\\
&\quad
-
\frac{\partial V_M}{\partial a},
\\
\left(-\nabla^2+m_\omega^2\right)\omega(\mathbf{x})
&=
g_{\omega NN}
N^\dagger(\mathbf{x})N(\mathbf{x})
\nonumber\\
&\quad
-
2\lambda_{\omega\rho}
g_{\omega NN}^2
g_{\rho NN}^2
\omega\rho^2,
\\
\left(-\nabla^2+m_\rho^2\right)\rho(\mathbf{x})
&=
g_{\rho NN}
N^\dagger(\mathbf{x})
\frac{\tau_3}{2}
N(\mathbf{x})
\nonumber\\
&\quad
-
2\lambda_{\omega\rho}
g_{\omega NN}^2
g_{\rho NN}^2
\omega^2\rho,
\\
-\nabla^2 A_0(\mathbf{x})
&=
e\,
N^\dagger(\mathbf{x})
\frac{1+\tau_3}{2}
N(\mathbf{x}),
\label{eq:meson_eom}
\end{align}
for the mesons and 
\begin{equation}
\big[ \vec{\alpha} \cdot \vec{p} + \beta\, m^*_N +V \big]N_i = \epsilon_i N_i\ , \label{nucleon}
\end{equation}
for the nucleon.

\section{\label{sec:level_i}parameter determination}

In this work, relativistic Hartree calculations in coordinate space are performed for spherical nuclei with closed or nearly closed shells.

In the present model, we have 12 parameters to determine:
\begin{equation}
m_0,\,
g_1,\,
g_2,\,
\bar{\mu}_\sigma^2,\,
\lambda_4,\,
\lambda_6,\,
g_\omega,\,
\bar{\mu}_a^2,\,
\gamma_4,\,
g_\rho,\,
\lambda_{\omega\rho},\,
\lambda_6'.
\end{equation}
We want to find the parameters such that the models reproduce as good as possible the ground state properties of the nuclei as well as the saturation properties of the uniform nuclear matter. Therefore, we optimize the parameters of PDM to the binding energies and charge radii for the selected nuclei, and the saturation properties of the nuclear matter.

In the fitting procedure of $a_0$ model, we take $m_0$, $\lambda_6$, $g_\rho$, $\lambda_{\omega\rho}$, $\lambda_6'$, $\lambda_4$, $g_\omega$ as independent fitting variables and determine their optimal values by minimizing the $\chi^2$ 
function,
\begin{align}
\chi^{2} &=
\sum_{i}
\left(
  \frac{E^{\rm calc}_{i}/A - E^{\rm exp}_{i}/A}{0.05~\text{MeV}}
\right)^{2} \nonumber
+
\sum_{i}
\left(
  \frac{r^{\rm calc}_{c,i} - r^{\rm exp}_{c,i}}{0.025~\text{fm}}
\right)^{2} \nonumber
\\[6pt]
&\quad+
\left(
  \frac{K_{0} - 240~\text{MeV}}{20~\text{MeV}}
\right)^{2} \nonumber
+
\left(
  \frac{S_{0} - 31.7~\text{MeV}}{1.1~\text{MeV}}
\right)^{2}
\\[6pt]
&\quad+
\left(
  \frac{n_{0} - 0.150~\text{fm}^{-3}}{0.010~\text{fm}^{-3}}
\right)^{2}
+
\left(
  \frac{B_{0} - 16~\text{MeV}}{1~\text{MeV}}
\right)^{2} \ ,
\end{align}
where the index $i$ 
runs over the
16 nuclei with various mass 
numbers:
$^{16}$O, $^{40}$Ca, $^{90}$Zr, $^{114}$Sn, $^{116}$Sn,
$^{118}$Sn, $^{120}$Sn, $^{122}$Sn, $^{124}$Sn, $^{126}$Sn,
$^{128}$Sn, $^{130}$Sn, $^{132}$Sn, $^{140}$Ce,
$^{208}$Pb, and $^{214}$Pb.
The quantity $n_0$ denotes the saturation density, and $B_0$, $K_0$, and $S_0$ the binding energy per nucleon, incompressibility, symmetry energy at saturation, respectively.

For each trial set of the independent fitting variables, the remaining 5 parameters are determined as follows. 
$g_1$ and $g_2$ are determined from the vacuum masses of positive and negative parity nucleons in Eq.~(\ref{maj}), where we identify $N^+$ as $N(939)$ and $N^-$ as $N(1535)$. 
$\bar{\mu}_\sigma^2$ is determined from the stationary condition in the vacuum:
\begin{equation}
-\bar{\mu}_\sigma^2
+\lambda_4 f_\pi^2
-\lambda_6 f_\pi^4
-m_\pi^2
=0.
\end{equation}
The $\eta$ meson mass in the vacuum is expressed as 
\begin{equation}
m_\eta^2 = \bar{\mu}_\sigma^2 - \bar{\mu}_a^2 + m_\pi^2 \ ,
\end{equation}
so that the value of $\mu_a^2$ is determined from the given value of $\bar{\mu}_\sigma^2$ using the meson masses listed in Table~\ref{vacMT1}.
Finally, the value of $\gamma_4$ is determined as
\begin{align}
\gamma_4 = & \quad \frac{m_{a_0}^2 + (5 \lambda_{6} - 2\lambda_{6}') f_{\pi}^4 + \bar{\mu}^2_{a}}{f_{\pi}^2} \ .
\end{align}
The same fitting method is employed for the model without $a_0$ meson except that $\lambda_6'$, $\bar{\mu}^2_a$ and $\gamma_4$ do not exist in the model without $a_0$.
\begin{table}[h!]
\caption{\label{vacMT1}Vacuum mass of the mesons.}
\begin{ruledtabular}
\begin{tabular}{ccccc}
\textrm{$\pi$[MeV]}&
\textrm{$a_0$(980)[MeV]}&
\textrm{$\eta$[MeV]}&
\textrm{$\omega$[MeV]}&
\textrm{$\rho$[MeV]}\\
\colrule
138 & 980 & 550 & 783 & 776\\
\end{tabular}
\end{ruledtabular}
\end{table} 

The experimental binding energies are obtained from the AME2020~\cite{Wang_2021}, while the experimental charge radii are taken from~\cite{ANGELI201369}. We note that the slope parameter $L_0$ is not included in the fitting procedure because it is an inferred nuclear matter quantity whose empirical determination remains model dependent.
Instead, we include binding energies per nucleon and charge radii of selected Sn isotopes in the fit to help constrain the isovector sector through their dependence on neutron-proton asymmetry. The value of $L_0$ is therefore regarded as a prediction.

The best fitted parameters for PDM with $a_0$ and PDM without $a_0$ are listed in Table~\ref{tab:fit_parameters}. 
\begin{table}[h]
\caption{\label{tab:fit_parameters}
Best-fit parameters of the PDM with and without the $a_0$ meson.
The quoted uncertainties correspond to $1\sigma$ statistical uncertainties.}
\begin{ruledtabular}
\begin{tabular}{ccc}
Parameter & With $a_0$ & Without $a_0$ \\
\hline
$m_0$ (MeV)
    & $689 \pm 71$
    & $689 \pm 69$ \\

$g_1$
    & $7.90 \pm 0.52$
    & $7.89 \pm 0.50$ \\

$g_2$
    & $14.35 \pm 0.52$
    & $14.34 \pm 0.50$ \\

$\bar{\mu}_\sigma^2/f_\pi^2$
    & $20.80 \pm 3.65$
    & $20.77 \pm 3.55$ \\

$\lambda_4$
    & $38.31 \pm 5.84$
    & $38.26 \pm 5.67$ \\

$\lambda_6 f_\pi^2$
    & $15.28 \pm 2.19$
    & $15.26 \pm 2.13$ \\

$g_\omega$
    & $7.39 \pm 1.15$
    & $7.38 \pm 1.12$ \\

$\bar{\mu}_a^2/f_\pi^2$
    & $-12.40 \pm 3.65$
    & $-$ \\

$\gamma_4$
    & $280.70 \pm 181.51$
    & $-$ \\

$g_\rho$
    & $11.05 \pm 1.91$
    & $9.44 \pm 0.94$ \\

$\lambda_{\omega\rho}$
    & $0.0077 \pm 0.0378$
    & $0.0776 \pm 0.0657$ \\

$\lambda_6'/\lambda_6$
    & $-3.41 \pm 6.00$
    & $-$ \\
\hline
$\chi^2$
    & $26.49$
    & $26.51$ \\

d.o.f.
    & $29$
    & $30$ \\

$\chi^2/\mathrm{d.o.f.}$
    & $0.91$
    & $0.88$ \\
\end{tabular}
\end{ruledtabular}
\end{table}
This table shows that both models reproduce the inputs very well with the reduced $\chi^2 \sim 1$.
Our fits yield $m_0=689 \pm 71$ MeV for the model with the $a_0$ meson and $689 \pm 69$ MeV for the no-$a_0$ model.
Both values fall within the broad range $m_0\sim400–900$ MeV obtained in previous studies employing different frameworks and observables, including analyses of nuclear matter and NS properties~\cite{PhysRevC.106.065205, Minamikawa:2023key, PhysRevC.103.045205, PhysRevC.108.055206, SciChinaPhysMechAstron6920263232011, PhysRevC.109.065807, nlhh-xqjp, xwkv-s8lg, universe11100345}, finite nuclei~\cite{YKimNSPDM2023}, gravitational form factors~\cite{KAWAGUCHI2026140400}, and the electric form factor of proton~\cite{Harada:2026dyj}.
In particular, in Ref.~\cite{PhysRevC.108.055206}, 
the same model is used to perform the analysis of neutron stars and obtained the constraint 
$580 \,\text{ MeV} \lesssim m_0 \lesssim 860 \,\text{ MeV} $ for the $a_0$ model and $ 540 \lesssim m_0 \lesssim 870$ MeV for no-$a_0$ model, for $L_0=60$ MeV. 
The value of $m_0$ obtained in this paper lies well inside these intervals and selects a much narrower region around $m_0\sim 700$ MeV. The result of no-$a_0$ model is also consistent with the one obtained in Ref.~\cite{YKimNSPDM2023}. This indicates that finite nucleus observables provide complementary information that can substantially restrict the parameter region.

\section{\label{sec:SP}Nuclear matter properties}
In this section, we present the nuclear matter properties calculated from the current models focusing on the effect of $a_0$ meson.

In Table~\ref{SPT1}, the nuclear matter properties, including the saturation density $n_0$, binding energy per nucleon $B_0$, incompressibility $K_0$, symmetry energy $S_0$, slope parameter $L_0$, symmetry incompressibility $K_{sym}$, and symmetry skewness $Q_{sym}$ are listed for the PDMs with and without $a_0$ meson, with comparison to the meson exchange model NL3$^*$~\cite{LALAZISSIS200936} and point coupling model PC-PK1~\cite{PhysRevC.82.054319}. Both models reproduce the inputs very well.
\begin{table*}
\centering
\caption{
Saturation properties calculated from the best-fit parameters in the PDM
with and without the $a_0$ meson, compared with other models.
The quoted uncertainties correspond to $1\sigma$ uncertainties.
\label{SPT1}}
\begin{ruledtabular}
\begin{tabular}{lcccc}
\textbf{Property}
& \textbf{$a_0$-PDM}
& \textbf{no-$a_0$ PDM}
& \textbf{NL3$^*$~\cite{LALAZISSIS200936}}
& \textbf{PC-PK1~\cite{PhysRevC.82.054319}} \\
\colrule

$n_0$ (fm$^{-3}$)
& $0.161 \pm 0.002$
& $0.161 \pm 0.002$
& 0.150
& 0.154 \\

$B_0$ (MeV)
& $16.63 \pm 0.09$
& $16.64 \pm 0.09$
& 16.31
& 16.12 \\

$K_0$ (MeV)
& $234 \pm 6$
& $234 \pm 6$
& 258
& 238 \\

$S_0$ (MeV)
& $32.1 \pm 1.0$
& $32.2 \pm 1.0$
& 38.7
& 35.6 \\

\colrule

$L_0$ (MeV)
& $100.7 \pm 30.4$
& $63.3 \pm 20.1$
& $122.6$~\cite{PhysRevC.82.054314}
& 113 \\

$K_{\rm sym}$ (MeV)
& $60 \pm 225$
& $-169 \pm 68$
& $-630$~\cite{PhysRevC.82.054319}
& $-583$ \\

$Q_{\rm sym}$ (MeV)
& $118 \pm 528$
& $226 \pm 694$
& $-$
& $-$ \\

\end{tabular}
\end{ruledtabular}
\end{table*}

The current models predict $L_0 = 100.7 \pm 30.4$ MeV for $a_0$ model and $L_0 = 63.3 \pm 20.1$ MeV for no-$a_0$ model. Still, the predicted $L_0$ agrees with the value summarized in Ref.~\cite{universe7060182} as $L_0 \approx 57.7 \pm 19$ MeV from the theoretical neutron star calculations and recent constraint $42 < L_0  < 117$ MeV from the study of charged pion spectra at high transverse momenta due to the S$\pi$RIT collaboration~\cite{PhysRevLett.126.162701}. 
We note that the inclusion of the $a_0$ meson shifts the central value of $L_0$ by about $40$\,MeV because the inclusion of $a_0$ meson increases the $\rho$ meson coupling $g_{\rho NN}$ to reproduce the symmetry energy at saturation.

We further examine the symmetry incompressibility and skewness parameters, $K_{sym}$ and $Q_{sym}$, evaluated at saturation.
The PDM with and without the $a_0$ meson predicts $K_{sym}=60\pm225$ MeV and $K_{sym}=-169\pm68$ MeV, respectively.
Including the $a_0$ meson in the model shifts the central value of $K_{\rm sym}$ upward. Both results are compatible within their uncertainties with the estimate $K_{sym}\approx-107\pm88$ MeV summarized from neutron-star analyses in Ref.~\cite{universe7060182}. For the skewness parameter, we obtain $Q_{sym}=118\pm528$ MeV with $a_0$ and $Q_{sym}=226\pm694$ MeV without $a_0$.
The uncertainties are substantially larger than the difference between the central values, leaving the sensitivity of $Q_{sym}$ to the $a_0$ effect unresolved.

\begin{figure}[t]
\includegraphics[width=\columnwidth]{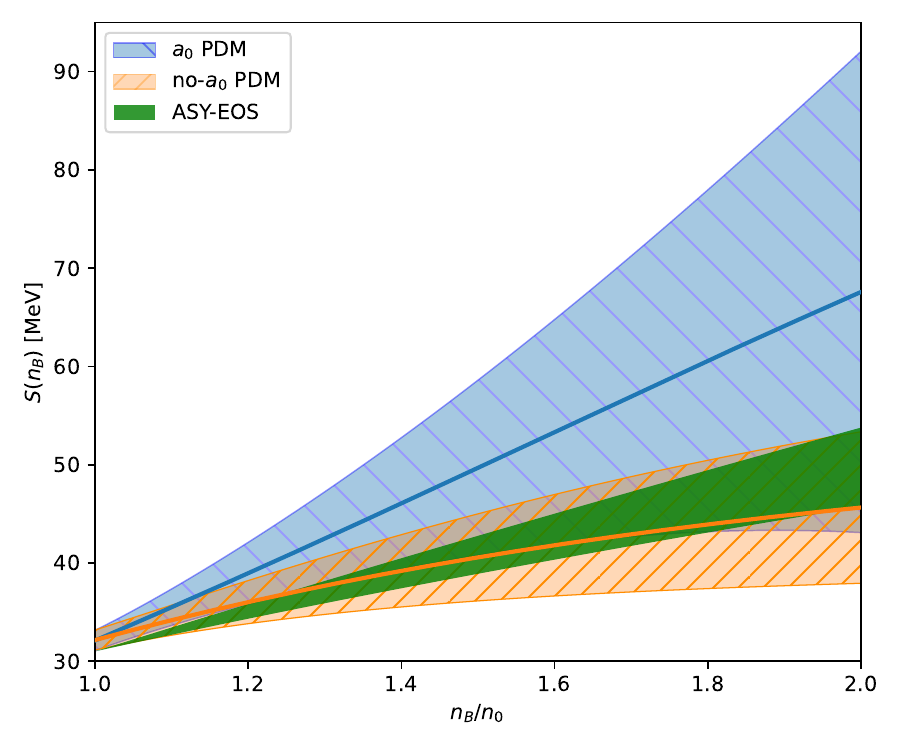}
\caption{\label{fig:S_density}Symmetry energy $S(n_B)$ for PDM with and without a0. 
The result obtained from the ASY-EOS experiment~\cite{PhysRevC.94.034608} is shown by the green band.
}
\end{figure}

Figure~\ref{fig:S_density} shows the density dependence of the symmetry energy $S(n_B)$ for PDM with and without $a_0$ meson. We note that both models agree well with the analysis from ASY-EOS experiment~\cite{PhysRevC.94.034608} in the case of $S(n_0) = 31$\,MeV, which is parametrized as $S(n_B) = 19\text{MeV}(n_B/n_0)^\gamma + 12\text{MeV}(n_B/n_0)^{2/3}$ with $\gamma = 0.68 \pm 0.19$ (grey band). The central value of the symmetry energy predicted from the $a_0$ model is enhanced compared with the one from no-$a_0$ model, and the difference between them is larger for larger density. However, the symmetry energy in both models still consistent with each other within the uncertainty, and we cannot distinguish clearly the effect of $a_0$. The constraints on $K_{sym}$ and $Q_{sym}$, as well as $L_0$ from future experiments may help us to clarify the effect of $a_0$ meson.

\section{\label{sec:NP}nuclear properties}

In this section, we present some properties of nuclei calculated from the PDMs.

Tables~\ref{tab:EA_selected} and \ref{tab:rc_selected} show the binding energies per nucleon $E/A$ and charge radii $r_c$ for the fitted nuclei computed from the PDM with and without $a_0$ meson together with the experimental data for binding energies from AME2020~\cite{Wang_2021} and charge radii from~\cite{ANGELI201369}. 
For the selected nuclei, the root-mean-square deviations (RMSDs) of the $a_0$-PDM 
defined by
\begin{align}
\mbox{RMSD}_{E/A} & =\sqrt{\frac{1}{N}\sum_{i=1}^{N}
\left[(E/A)_i^{\mathrm{theo}}-(E/A)_i^{\mathrm{exp}}\right]^2} \ ,\notag\\
\mbox{RMSD}_{r_c} & = \sqrt{\frac{1}{N}\sum_{i=1}^{N}
\left(r_{c,i}^{\mathrm{theo}}-r_{c,i}^{\mathrm{exp}}\right)^2} \ ,
\end{align}
are $0.048$\,MeV for $E/A$ and $0.020$\,fm for $r_c$. For the PDM without $a_0$ meson, the corresponding RMSDs are $0.049$\,MeV and $0.019$\,fm, respectively. Thus, including the $a_0$ meson gives a slightly better RMSD for the binding energy per nucleon, but a slightly larger RMSD for the charge radius. 
Since these differences are very small, we conclude that the $a_0$ meson has only a negligible effect on the overall description of binding energies and charge radii. 
The comparable quality of the results obtained with and without the $a_0$ meson further suggests that binding energies and charge radii, being predominantly bulk isoscalar observables, provide only limited sensitivity to the different treatments of the isovector sector.
\begin{table}
\scriptsize
\caption{\label{tab:EA_selected}
Binding energy per nucleon $E/A$ for the fitted nuclei with fitting
uncertainties. All values are in MeV.
The $\mbox{RMSD}_{E/A}=\sqrt{\frac{1}{N}\sum_{i=1}^{N}
\left[(E/A)_i^{\mathrm{theo}}-(E/A)_i^{\mathrm{exp}}\right]^2}$
are computed using the best-fit parameters.
The experimental data are retrieved from AME2020~\cite{Wang_2021}.
}
\begin{ruledtabular}
\begin{tabular}{cccc}
Nucleus & Experiment & With $a_0$ & Without $a_0$ \\
\hline
$^{16}$O
& $7.976$ & $7.989 \pm 0.042$ & $7.995 \pm 0.041$ \\

$^{40}$Ca
& $8.551$ & $8.604 \pm 0.025$ & $8.612 \pm 0.024$ \\

$^{90}$Zr
& $8.710$ & $8.617 \pm 0.027$ & $8.620 \pm 0.019$ \\

$^{116}$Sn
& $8.523$ & $8.496 \pm 0.019$ & $8.503 \pm 0.018$ \\

$^{124}$Sn
& $8.467$ & $8.456 \pm 0.014$ & $8.456 \pm 0.012$ \\

$^{132}$Sn
& $8.355$ & $8.322 \pm 0.029$ & $8.309 \pm 0.025$ \\

$^{208}$Pb
& $7.867$ & $7.922 \pm 0.018$ & $7.923 \pm 0.016$ \\

$^{214}$Pb
& $7.772$ & $7.884 \pm 0.025$ & $7.880 \pm 0.023$ \\

$^{114}$Sn
& $8.523$ & $8.467 \pm 0.021$ & $8.475 \pm 0.020$ \\

$^{140}$Ce
& $8.376$ & $8.394 \pm 0.015$ & $8.395 \pm 0.015$ \\

$^{118}$Sn
& $8.517$ & $8.522 \pm 0.017$ & $8.529 \pm 0.016$ \\

$^{120}$Sn
& $8.504$ & $8.539 \pm 0.015$ & $8.545 \pm 0.015$ \\

$^{122}$Sn
& $8.488$ & $8.496 \pm 0.014$ & $8.500 \pm 0.013$ \\

$^{126}$Sn
& $8.444$ & $8.418 \pm 0.016$ & $8.416 \pm 0.014$ \\

$^{128}$Sn
& $8.417$ & $8.384 \pm 0.019$ & $8.378 \pm 0.017$ \\

$^{130}$Sn
& $8.387$ & $8.351 \pm 0.024$ & $8.342 \pm 0.021$ \\
\hline
RMSD
& --- & $0.048$ & $0.049$ \\
\end{tabular}
\end{ruledtabular}
\end{table}
\begin{table}
\scriptsize
\caption{\label{tab:rc_selected}
Charge radii $r_{\rm c}$ for the fitted nuclei with fitting uncertainties.
All values are in fm.
The $\mbox{RMSD}_{r_c}= \sqrt{\frac{1}{N}\sum_{i=1}^{N}
\left(r_{c,i}^{\mathrm{theo}}-r_{c,i}^{\mathrm{exp}}\right)^2}$
are computed using the best-fit parameters.
The experimental data are obtained from~\cite{ANGELI201369}.
}
\begin{ruledtabular}
\begin{tabular}{cccc}
Nucleus & Experiment & With $a_0$ & Without $a_0$ \\
\hline
$^{16}$O
& $2.699$ & $2.764 \pm 0.015$ & $2.758 \pm 0.013$ \\

$^{40}$Ca
& $3.478$ & $3.466 \pm 0.012$ & $3.462 \pm 0.011$ \\

$^{90}$Zr
& $4.269$ & $4.273 \pm 0.006$ & $4.273 \pm 0.006$ \\

$^{116}$Sn
& $4.625$ & $4.612 \pm 0.006$ & $4.614 \pm 0.006$ \\

$^{124}$Sn
& $4.674$ & $4.664 \pm 0.006$ & $4.666 \pm 0.006$ \\

$^{132}$Sn
& $4.709$ & $4.731 \pm 0.007$ & $4.732 \pm 0.007$ \\

$^{208}$Pb
& $5.501$ & $5.508 \pm 0.010$ & $5.508 \pm 0.010$ \\

$^{214}$Pb
& $5.558$ & $5.556 \pm 0.010$ & $5.555 \pm 0.010$ \\

$^{114}$Sn
& $4.610$ & $4.601 \pm 0.006$ & $4.602 \pm 0.006$ \\

$^{140}$Ce
& $4.877$ & $4.880 \pm 0.006$ & $4.880 \pm 0.006$ \\

$^{118}$Sn
& $4.639$ & $4.624 \pm 0.006$ & $4.625 \pm 0.006$ \\

$^{120}$Sn
& $4.652$ & $4.633 \pm 0.006$ & $4.635 \pm 0.006$ \\

$^{122}$Sn
& $4.663$ & $4.648 \pm 0.006$ & $4.650 \pm 0.006$ \\

$^{126}$Sn
& $4.683$ & $4.681 \pm 0.006$ & $4.682 \pm 0.006$ \\

$^{128}$Sn
& $4.692$ & $4.697 \pm 0.006$ & $4.699 \pm 0.006$ \\

$^{130}$Sn
& $4.702$ & $4.714 \pm 0.007$ & $4.715 \pm 0.007$ \\
\hline
RMSD
& --- & $0.020$ & $0.019$ \\
\end{tabular}
\end{ruledtabular}
\end{table}

\begin{figure}[t]
\includegraphics[width=\columnwidth]{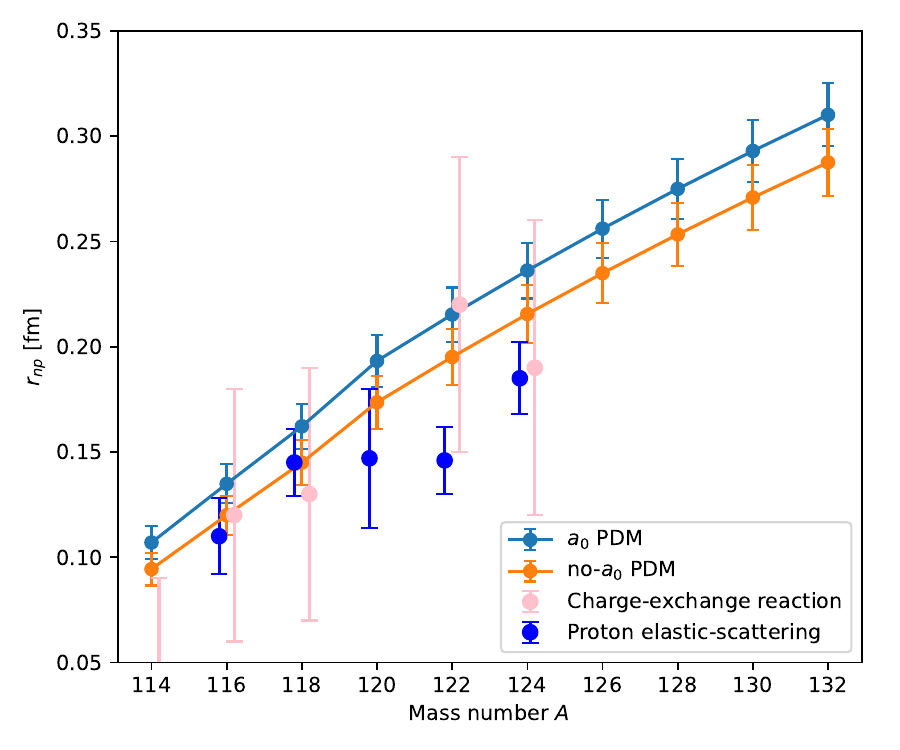}
\caption{\label{fig:rnp_sn}Neutron-skin thicknesses of the even-A Sn isotopes from $^{114}$Sn to $^{132}$Sn calculated from the current models.
The pink points with error bars denote values extracted from
charge-exchange reactions for $^{114,116,118,122,124}$Sn
\cite{PhysRevLett.82.3216}, while the blue points with error bars
show results obtained from proton elastic-scattering analyses
for $^{116,118,120,122,124}$Sn \cite{PhysRevC.77.024317}.
}
\end{figure}

Neutron skin thickness $r_{np} \equiv r_n -r_p$ serves as a sensitive and experimentally accessible probe for the isovector properties of nuclear matter.
Furthermore, it is strongly correlated with the density dependence of the equation of state for isospin-asymmetric matter, offering a more suitable observable for elucidating the $a_0$ meson effect. 
It is also important for the study of the asymmetric matter properties such as EoS
because neutron skin thickness is closely related to the density dependence of the symmetry energy.
Figure~\ref{fig:rnp_sn} shows the values of the neutron skin thicknesses of the even-even Sn isotopes calculated in the $a_0$ PDM (shown by muted blue line) and no-$a_0$ PDM (by orange line), together with values extracted from charge-exchange reaction~\cite{PhysRevLett.82.3216} (by pick dots) and proton elastic-scattering measurements~\cite{PhysRevC.77.024317} (by blue dots). 
Both models capture the overall magnitude and increasing trend of $r_{np}$ with neutron excess. The best-fit $a_0$ PDM consistently predicts larger neutron skin thicknesses than the best-fit no-$a_0$ PDM. However, the overlapping statistical uncertainty bands indicate that the quoted uncertainties alone do not establish a clear separation between the two predictions. A similar upward shift in the central predictions due to the inclusion of $a_0$ meson is also found for the other nuclei examined, suggesting a systematic tendency within the present fits.

\begin{figure}[t]
\includegraphics[width=\columnwidth]{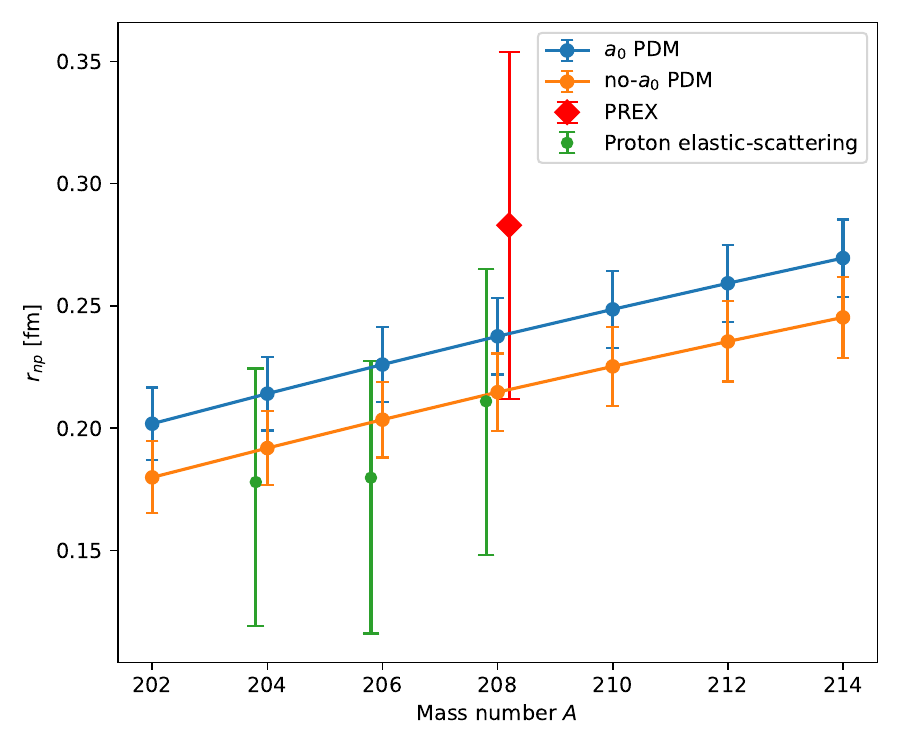}
\caption{\label{fig:rnp_pb}
Neutron skin thickness of the even-A Pb isotopes from $^{202}$Pb to $^{214}$Pb
calculated from the current models.
The green points with error bar show the results for $^{204,206,208}$Pb extracted from proton elastic scattering~\cite{PhysRevC.82.044611}, and the red point with error bar denotes the neutron skin thickness for $^{208}$Pb reported by PREX collaboration~\cite{PhysRevLett.126.172502}.
}
\end{figure}

The $r_{np}$ of the even-even Pb isotopic chain are shown in Fig.~\ref{fig:rnp_pb}, compared with 
the results for $^{204,206,208}$Pb from proton elastic scattering~\cite{PhysRevC.82.044611} and $r_{np} = 0.283 \pm 0.071$\,fm for 
 $^{208}$Pb
extracted from parity-violating electron scattering by the PREX collaboration~\cite{PhysRevLett.126.172502}.
Both models predict an increase in $r_{np}$ with increasing neutron excess, with the best-fit $a_0$ model yielding systematically larger neutron skin thicknesses. For $^{208}$Pb, the $a_0$ and no-$a_0$ PDMs predict $r_{np} = 0.238 \pm 0.016   $ fm and $r_{np} = 0.215 \pm 0.016 $ fm, respectively. The central predictions of both models lie within the experimental $1\sigma$ intervals reported in Refs.~\cite{PhysRevLett.126.172502} and~\cite{PhysRevC.82.044611}. The prediction of $a_0$ model lies closer to the PREX central value, although this comparison alone does not establish a preference for either model and we cannot resolve the effect of $a_0$ meson within the fitting uncertainty.
We would like to stress that we do not include neutron skin thickness of $^{208}\text{Pb}$, rather, we give it as a prediction. 

For $^{48}\text{Ca}$, the $a_0$ and no-$a_0$ PDMs predict $r_{np} = 0.227 \pm 0.007 $ fm and $r_{np} = 0.220 \pm 0.007 $ fm, respectively. 
The values are much larger than the measured value $0.121 \pm  0.026$ fm obtained from the parity-violating elastic electron scattering by the CREX collaboration~\cite{PhysRevLett.129.042501}.

\section{\label{sec:NM}summary and discussion}

We studied the nuclear structure using the PDM emphasizing the effect of isovector scalar meson $a_0(980)$.  
An extensive search of the PDM parameters for models with and without the $a_0$ was performed by fitting the model parameters 
to the binding energies and charge radii of the selected nuclei listed in 
Tables~\ref{tab:EA_selected} and~\ref{tab:rc_selected},
together with 
the nuclear matter saturation properties: the saturation density $n_0$, the binding energy per nucleon $B_0$, incompressibility $K_0$, and symmetry energy at saturation $S_0$. Both models reproduce the inputs well, with reduced $\chi^2$ values close to unity.

The fits constrain the chiral invariant mass of nucleon to $689 \pm 71$ MeV
for $a_0$ model and $689 \pm 69$ MeV
for no-$a_0$ model, both of which are consistent with constraint obtained from the studies on neutron star analyses  (See e.g. Ref.~\cite{sym16091238} and references therein.) and the nucleon gravitational form factor~\cite{KAWAGUCHI2026140400}.

We then investigate the effect of the $a_0$ meson to the slope $L_0$, curvature $K_{sym}$, and skewness $Q_{sym}$ at saturation density, and the density dependence of symmetry energy $S(n_B)$.
The current models predict $L_0 =100.7 \pm 30.4$ MeV for $a_0$ model and $L_0 = 63.3 \pm 20.1$ MeV for no-$a_0$ model, which implies that the inclusion of $a_0$ meson enlarge the central value of $L_0$ by about 40 MeV. 
We note that the predictions agree with the analyses from neutron star calculations~\cite{universe7060182} and charged pion spectra due to S$\pi$RIT collaboration~\cite{PhysRevLett.126.162701}.

The inclusion of the $a_0$ meson also raises the central value of $K_{\mathrm{sym}}$, but the predictions with and without $a_0$ overlap within their uncertainties, preventing a clear distinction between the two models. The fitting uncertainties for $Q_{\mathrm{sym}}$ are also too large for us to clearly distinguish the effect of $a_0$ meson. In addition, $a_0$ meson enhances the central prediction for the symmetry energy $S(n_B)$ at suprasaturation densities, with the difference increasing with density. Both models are consistent with the constraints from ASY-EOS~\cite{PhysRevC.94.034608}. Future experimental constraints on the density dependence of the symmetry energy could help to further understand the role of the $a_0$ meson in the equation of state of isospin-asymmetric matter.

We showed that the binding energies and the charge radii of selected nuclei are well reproduced by both the $a_0$ PDM and no-$a_0$ PDM: the RMSD of binding energy per nucleon are in the order of $0.05$\,MeV for the fitted nuclei; the RMSDs of charge radii are in the order of $0.02$\,fm for the fitted nuclei.
Our results showed that the inclusion of $a_0$ meson has only negligible effect to the binding energies and charge radii.

The neutron skin thickness of even-even Sn and Pb isotopes was studied and compared to the experiments.
We found that the inclusion of $a_0$ meson enhances the central value of neutron skin thickness across different isotopes systematically. 
However, since the error bars are large, the predictions from both the $a_0$ and no-$a_0$ models are consistent with each other.

The neutron skin thickness of $^{208}\mathrm{Pb}$ are predicted to be $r_{np} = 0.238 \pm 0.016   $ fm
for $a_0$ PDM and 
$r_{np} = 0.215 \pm 0.016 $ fm
for no-$a_0$ PDM.
The predictions are 
consistent with the PREX measurement $r_{np} = 0.283 \pm 0.071$\,fm within 1\,$\sigma$ uncertainty. 
On the other hand, the neutron skin thickness of $^{48}\mathrm{Ca}$ are calculated to be $r_{np} = 0.227 \pm 0.007 $ fm for $a_0$ model and $r_{np} = 0.220 \pm 0.007 $ fm for no-$a_0$ model.
Both the values are large compared to the CREX measurement $r_{np} = 0.121 \pm  0.026$\,fm.

In future, it is interesting to investigate the role of isovector tensor mesons in the PDM. 
Recent studies have shown that an enhanced isovector spin-orbit interaction can reduce the predicted neutron skin of $^{48}\mathrm{Ca}$ while having a comparatively small effect on that of $^{208}\mathrm{Pb}$, improving the simultaneous description of the PREX and CREX charge weak form-factor data~\cite{YUE20261270,QIU2026140648}. Since an enhanced isovector tensor coupling can generate such a spin-orbit interaction in a covariant framework~\cite{QIU2026140648}, incorporating this coupling into the PDM would allow us to examine whether this mechanism can alleviate the PREX-CREX tension within the present framework.

The present study will be also  extended to more neutron-rich nuclei near the neutron drip line, where the large neutron-proton asymmetry may enhance the sensitivity to the isovector sector. In such weakly bound systems, continuum and pairing effects become increasingly important, and their interplay with the isovector interactions deserves further investigation.

\begin{acknowledgments}
The work of  M.H. is supported in part by JSPS KAKENHI Grant Nos.~23H05439 and 24K07045. 
 This work was supported in part by the Institute for Basic Science (IBS-R031-D1) and by the National Research Foundation 
of Korea funded by the Ministry of Science and ICT (RS-2024-00436392).
Y. K. Kong would like to take this opportunity to thank the ``THERS Make New Standards Program for the Next Generation Researchers" and Kobayashi-Maskawa Institute for the Origin of Particles and the Universe (KMI), Nagoya University through the ``Overseas Dispatch Program for Young Researchers".
\end{acknowledgments}

\bibliography{apssamp}



\end{document}